\documentstyle[12pt,fleqn,espcrc1,epsf]{article}

\title{QCD and Nuclear Physics\thanks{Plenary talk given at the International Nuclear 
Physics
Conference, August 24-28, 1998 in Paris},\thanks{This work is supported in part by the 
Department
of Energy, Grant DE-FG02-94ER40819}}%

\author{A.H. Mueller\address{Department of Physics, 
        Columbia University\\
 New York, New York 10027, USA}}

\begin{document}
\maketitle
\begin{abstract}
The main part of this talk is a review and summary of how QCD is used in two main areas 
of nuclear
physics, namely in determining the quark flavor and spin content of the proton and in ultrarelativistic heavy ion collisions.  Brief comments are made concerning effective 
theories in
hadron physics and on the separation of various twists in using the operator product 
expansion to
analyze hard processes.

\end{abstract}

\section{Introduction}

Quantum Chromodynamics (QCD) is a fundamental theory in the sense that one expects 
QCD to exist
not just as a perturbative expansion but also in its strong coupling regime.  There is good
evidence from lattice QCD calculations that this is indeed the case.  QCD stands in contrast 
to
QED or the Electroweak Theory which have good perturbative expansions but which are 
not expected
to exist in the strong coupling regime.  In particular it is widely believed that the 
Electroweak
Theory must be unified with additional interactions at a scale below its Landau ghost.

The likelihood that only asymptotically free theories make sense puts severe restrictions on 
the
use of ``effective'' field theories, field theories that are to be used in a limited range of
scales.  For example, it probably does not make much sense to try and describe low energy
pion-nucleon interactions in terms of a pseudoscalar meson-nucleon interaction Lagrangian 
since
the coupling necessary in such a description is so strong that the theory is internally
inconsistent even at low energy scales.  Of course the Born terms (tree graphs) of such a 
theory
still make sense since they represent an analytic structure which is completely determined 
by the
lowest lying states of the pion-nucleon system.

There is a special property of pion interactions which makes it possible to use particular
effective theories as a description of low energy pion-nucleon interactions. That special 
property
is chiral symmetry.  One of the properties of the pion, as the Nambu-Goldstone boson of 
chiral
symmetry breaking, is the fact that its couplings become weak at low momentum. Thus if 
one uses an
effective chiral theory to describe low energy pion-nucleon interactions, and if a cutoff is
introduced so that the couplings never grows  large, then one can use that effective theory, 
even
in a nonperturbative manner, to make precise calculations\cite{rev,Wei,Ord,Par,Kap}.  
Such theories
tend to have quite a few parameters since the cutoff is not very high and all the dynamics 
above
the cutoff must be put into parameters of the effective theory. There has been much
interest\cite{rev,Sav} and progress in the last few years using effective theories to describe
light nuclei in terms of pion-pion and pion-nucleon interactions.  It is perhaps a bit too early
to decide whether this description is superior to the traditional potential approach, but one 
can
hope to get better insight as to where it is more useful to use quarks and gluons to describe
hadronic physics and where it is better to use effective degrees of freedom.

As far as is known QCD-type theories are the only four-dimensional field theories 
``known'' to
exist as fundamental theories.   The goal in strong interaction physics is to understand how 
QCD
works.  Few physicists doubt the validity of QCD as the theory of the strong interactions. 
However, QCD is a rich and sophisticated theory about which there is still much to 
understand and
appreciate.

I shall cover three topics in this talk.  Firstly, I shall review progress in understanding the
quark and gluon structure of the proton wavefunction.  This has become one of the most 
active
fields in nuclear physics in which the Jefferson Laboratory has now begun to make 
important new
contributions.  This part of my talk will have some overlap with and is complementary to 
the talks
of L. Cardman\cite{Car} and A. Magnon]\cite{Mag}.  The next topic to be covered is that 
of heavy
ion collisions and small-x physics.  The object here is to describe the relationships between 
the
physics being pursued in high energy small-x physics and that which will become available 
at heavy ion colliders at Brookhaven and at CERN.  The final topic is perhaps more of a remark, but 
it is a
remark which is important for medium energy physics.  The point is that the separation 
between
perturbative and nonperturbative QCD is not always so easy to define.  Nevertheless, one 
must
often make such a separation in medium energy phenomenology and care must be taken to 
insure that
that the separation, even if ``scheme'' dependent, is done consistently.  A good example is 
a recent
analysis by Kataev\cite{Kat} and his collaborators of the description of the $F_3$ structure
function in QCD.

\section{Spin and Strangeness in the proton}
\subsection{Spin and the constituent quark model}

In the constituent quark model the proton is made of two up-quarks and a down-quark.  In 
the
nonrelativistic  version the three quarks are in zero orbital angular momentum states and so 
the
spin of the proton is given by the sum of the spins of the quarks.  With the usual SU(6)
wavefunctions\cite{Iof}, and with $\Delta  q$ labeling the z-component of spin of the 
quark  $q$,
one has

\begin{equation}
\Delta u = 4/3, \Delta d = - 1/3
\end{equation}

\noindent leading to

\begin{equation}
G_A = \Delta u - \Delta d = 5/3
\end{equation}

\noindent and a total spin of the proton, $\Delta\Sigma$, given by

\begin{equation}
\Delta\Sigma = \Delta u + \Delta d = 1.
\end{equation}

\noindent The nonrelativistic quark model gives a reasonable picture of the proton but the 
value
of $G_A$ is clearly somewhat high.

\noindent Relativistic quark models give better agreement with experiment for $G_A.$ 
Typically\cite{Bro} one has

\begin{equation}
\Delta u \approx 1, \Delta d \approx - 1/4
\end{equation}

\noindent  giving

\begin{equation}
G_A=\Delta u - \Delta d \approx 5/4
\end{equation}

\noindent and

\begin{equation}
\Delta\Sigma = \Delta u + \Delta d \approx 3/4.
\end{equation}

\noindent In neither the nonrelativistic nor in the relativistic versions of the model is there
any room for strange quarks in the proton.  In addition to $G_A$ the quark model gives a 
very good
account of baryon magnetic moments and it has been the basis on which spectroscopy of 
mesons and
baryons has been discussed for some time\cite{Iof}.  Much of what follows will be 
concerned with
whether the quark model is able to give a reasonable account of the protons total spin.

\subsection {How to measure spin}

The axial vector current of flavor-f quarks is given by

\begin{equation}
j_{5\mu}^f = \tilde{q}_f\gamma_\mu\gamma_5 q_f.
\end{equation}

\noindent For free quarks

\begin{equation}
(ps\vert j_{5\mu}\vert ps) = 2m s_{\mu} \mathrel{\longrightarrow\atop \to \infty} 2\lambda
p_{\mu}
\end{equation}

\noindent where $s_\mu$ is the fermion spin four-vector corresponding to a quark with 
spin
orientation $\vec{s}$ in its rest system, and where $\lambda$ is the quark helicity.  In a 
frame
where quarks have a large longitudinal momentum it is convenient to quantize spin along 
the
direction of the large momentum.

In the quark-parton picture of the proton it is natural to suppose that

\begin{equation}
(P s\vert j_{5\mu}^f \vert Ps) = 2M_P s_\mu\Delta q_f
\end{equation}

\noindent where $\vert Ps >$ is a proton state, and where $\Delta q_f$ is the fraction of the
proton's spin carried by quarks of flavor  f.

In spin-dependent deep inelastic scattering on a proton one can measure a particular 
combinarning: Citation `Iof' on page 3 undefined on input line 179.
! Missintion
of axial vector current matrix elements, that given by

\begin{equation}
2Ms_\mu \int_0^1 dx\  g_1^P(x,Q^2) = {1\over 2} \sum_f e_f^2 (Ps\vert j_{5\mu}^f\vert
Ps).
\end{equation}

\noindent Scattering on neutrons gives an independent combination of axial vector currents.
Defining

\begin{equation}
\int_0^1 dx g_1^{P(N)}(x,Q^2) = \Gamma_1^{P(N)}(Q^2)
\end{equation}

\noindent and using (9), one finds

\begin{equation}
\Gamma_1^P = {1\over 2} ({4\over 9} \Delta u + {1\over 9} \Delta d + {1\over 9} \Delta
s)
\end{equation}

\noindent and

\begin{equation}
\Gamma_1^N = {1\over 2} ({4\over 9} \Delta d + {1\over 9} \Delta u + {1\over 9} \Delta
s).
\end{equation}

An additional relation, involving $\Delta u,\Delta d$ and $\Delta s$ comes from semi-
leptonic
hyperon decays\cite{Clo}.  After using SU(3) flavor symmetry one finds

\begin{equation}
3F - D = \Delta u + \Delta d - 2\Delta s.
\end{equation}

\noindent Equations (12) - (14) apparently (but see below) give enough information to 
determine
$\Delta q_f$ from  $\Gamma_1^P,\Gamma_1^N,F$ and $D.$

\subsection {Bare versus constituent quarks}

In the constituent quark model the proton is described in terms of quarks which are not 
really
point-like and which are not the quark degrees of freedom which appear in the QCD 
Lagrangian.  The
quark fields in (7) and (10) are the fundamental (bare) fields of the QCD Lagrangian.  A 
basic
assumption of the constituent quark model is that for static matrix elements one may replace 
the
fundamental quark fields by the constituent (effective) quark fields of the quark model.  
This
gives a method of calculating forward, or near forward, matrix elements of local bare quark
currents in terms of wavefunctions of the constituent quark model.  This is the basis on 
which
one can see if there is agreement between the quark model and results obtained from deep
inelastic scattering which do not directly measure constituent quarks but which, through 
relations
like (10), do lead to static matrix elements of local currents.

\subsection{A final subtlety}

While (9) is a plausible assumption in fact it is not quite right.  The correct relation is
\cite{Lam,Efr,Alt,Col}

\begin{equation}
(P s\vert j_{5\mu}^f\vert Ps) = 2 Ms_\mu(\Delta q_f - {\alpha\over 2\pi} \Delta G)
\end{equation}

\noindent with $\Delta G$ the amount of the proton's spin carried by gluons.  The 
argument for
(15) is subtle and still not without controversy.  I think the simplest argument for its 
validity
comes from considering an example in quantum electrodynamics with massless electrons.  
At order
$\alpha$ a virtual photon can break up into an electron-positron pair.  Because helicity is
conserved in electromagnetic interactions the helicities of the electron and positron will be 
of
opposite sign.  This if one takes the matrix element of $j_{5\mu} =
\bar{\psi}\gamma_\mu\gamma_5\psi$ between virtual photon states, at order $\alpha,$ one 
would
expect to get zero since $j_{5\mu},$ from (8), is supposed to measure electron (and 
positron)
helicities.  (The situation is illustrated in Fig.1.)  However, the matrix element turns out to 
be
$-{\alpha\over 2\pi},$ for a transversely polarized photon having helicity 1, despite the fact 
that
any given electron-positron state gives a zero value.  This is the axial anomaly and
$-{\alpha\over 2\pi}$ is the value of the axial anomaly.  In QCD exactly the same 
phenomenon
occurs.  The $-{\alpha\over 2\pi}$ in (15) is the anomaly value while $\Delta G$ gives the 
net
helicity of gluons in the proton.  The $- {\alpha\over 2\pi}$ comes from effects which 
cannot be
ascribed to the quark wavefunction of the proton and is one of the most profound elements 
in QCD.
\begin{figure}[htb]
\epsfxsize=3.5in
\epsfysize=2in
\epsfbox[0 0 349 138]{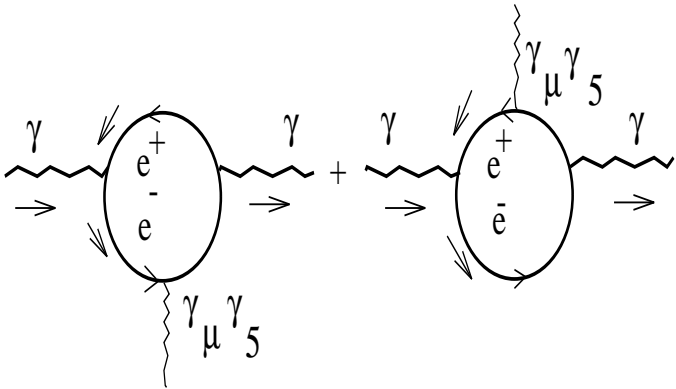}
\caption{The axial vector current couples to electrons and positrons having opposite 
helicities}
\label{Fig.1-928}
\end{figure}

\subsection{Results from spin-dependent deep inelastic scattering}

There are several groups that have done extensive global fits, using a second order
renormalization group formalism, to spin-dependent deep inelastic 
scattering\cite{Abe,SMC,Bal}. 
The experimental situation is discussed in some detail in the talk of A. Magnon\cite{Mag}.  
Here,
I shall briefly recount the results found by Altarelli et al who have carried out a series of
fits to all the available data.  The effect of polarized gluons is quite important in the various
fits.  The preferred fit has

\begin{eqnarray}
\Delta u = 0.86, \Delta d &=& - 0.37, \Delta s = - 0.05\\
\nonumber	\Delta G &=& 1.4 \pm 0.9
 \end{eqnarray}
 
 \noindent giving
 
\begin{equation}
 \Delta \Sigma = 0.44 \pm 0.09.
 \end{equation}

\noindent While these numbers are not in perfect agreement with the relativistic quark 
model the
situation is not so bad either. In particular, the reasonably small value for $\Delta s$ is 
much
more comfortable for the quark model that were early values.  However, because of the 
large errors
on $\Delta G$ it is perhaps premature to take the final results as conclusive.  The situation
would be considerably improved by a direct measurement of $\Delta G,$ as planned at 
COMPASS and
RHIC and which could also be done well at HERA.  A few reliable points for $\Delta 
G(x,Q^2)$ would
help considerably to fix the first moment, $\Delta G,$ of $\Delta G(x,Q^2).$

\subsection{Strangeness as viewed by vector currents}

In the previous sections we have seen how spin-dependent deep inelastic scattering gives
information on the polarized quark and gluon distributions of the proton.  New experiments 
at
Bates and the Jefferson Laboratory give complimentary information on the number of 
strange quarks
in the proton.  By measuring very accurately parity violating elastic electron-proton 
scattering
information on the contribution of strange quarks to the proton's electromagnetic form 
factors can
be obtained.  This is discussed in some detail in the talk of L. Cardman\cite{Car} so I will
simply state the results here and then comment on their significance and how they fit in 
with the
general program of determining the quark and gluon content of the proton.  If 
$G_{E(M)}^s(Q^2)$
stands for the strange quark contribution to the electric (magnetic) form factor of the proton 
at
a momentum transfer $Q^2$ then the new result from the SAMPLE experiment at Bates 
is\cite{Sam}

\begin{equation}
G_M^s(Q^2=0.1GeV^2) = 0.23\pm 0.37 \pm 0.15 \pm 0.19
\end{equation}

\noindent where the first error is statistical, the second is a systematic error and the third is
due to  the axial form factor $G_A^Z.$  The HAPPEX experiment at the
Jefferson Laboratory\cite{Hap} finds

\begin{equation}
G_E^s + 0.39\   G_M^s = 0.023\pm 0.034 \pm 0.022 \pm 0.026
\end{equation}

\noindent at $Q^2 = 0.47 GeV^2$ where the first error is statistical, the second systematic 
and
the third comes from uncertainties in the electric form factor of the neutron.  To set the scale
for these numbers we note that in the constituent quark model

\begin{equation}
G_M^u(0) = 2.47, G_M^d(0) = 0 \cdot 32
\end{equation}

\noindent and

\begin{equation}
G_E^u(0) = 4/3, G_E^d(0) = - 1/3
\end{equation}

\noindent give the up and down contributions to the proton's magnetic moment and charge,
respectively.

The HAPPEX result suggests that strange quarks give no more than a few percent of the up 
plus down
contribution to $G_E + 0.39 G_M.$  The SAMPLE result is also consistent with a small 
strange quark
contribution, but the limit is perhaps not too stringent.

\subsection{Summary on quark content of the proton}

Perhaps the main issue in determining the quark and gluon content of the proton is the issue 
of
how many and ``what kind'' of strange quarks are to be found in the proton's 
wavefunction.  At
first sight this would seem to be a simple problem whose answer, at least roughly, has 
been known
for some time. After all, spin-independent deep inelastic lepton-proton experiments have 
determined
that $x(s(x,Q^2) + \bar{s}(x,Q^2))$ is sizeable.  Indeed, the strange quark sea is about 
one-half that of the non-strange sea.

However, a moment's reflection is enough to realize that while it is interesting to know the
number, and the Bjorken\  x\  distribution, of strange quarks in the proton it is even more
important to know if those strange quarks are just short time fluctuations which play no 
role in the
dynamics of the proton or if the strange quarks live long enough to be essential in 
determining the
proton's mass and wavefunction.  For example, the fluctuation of a gluon of the proton 
into an
$s\bar{s}$ pair, in which the relative transverse momentum of the   $s$\ and $\bar{s}$  is 
greater
than a few GeV or so, contributes to the spin-independent strange sea distribution, but
such pairs are too compact and short-lived to interact with the rest of the proton and so are
uninteresting for static properties of the proton.  In particular such short-lived fluctuations
should not contribute to $\Delta s,$ because the \ $s$\ and $\bar{s}$ helicities will cancel 
out in
$\Delta s,$ and they should not contribute to $G_M^s$ or $G_E^s,$ again because the 
$\bar{s}$ and \
$s$\ will cancel.

We can now begin to appreciate the new information contained in spin-dependent deep 
inelastic
scattering and in strangeness as measured in parity violating elastic electron proton 
scattering. 
$\Delta s(Q^2)$ measures the sum of the $s$\  and\  $\bar{s}$\ helicity fractions in the 
proton
independently of the longitudinal momentum  fraction of the \  $s$\ and $\bar{s}$ and 
including
all transverse momenta up to  $Q.$  (One expects no $Q^2$-dependence of $\Delta s$ so 
long as $Q^2$
is grater than a few $GeV^2.$)  In order that the \ $s$\ and $\bar{s}$\ not have cancelling
helicities the transverse momentum of the\ $s$\ and \ $\bar{s}$\ must be small enough that 
they
have helicity nonconserving (nonperturbative) interactions in the proton or that the current 
quark
mass, about
$150 MeV,$ not be negligible.  For example, an interaction correlating the $\bar{s}$  with 
a
spectator u-quark, perhaps giving a virtual $K^+$ meson,  could certainly polarize the 
strange sea. 
Strange quarks measured by vector currents, as in the parity violating experiments, indicate 
a
difference between the transverse momentum (or transverse coordinate) distributions of the 
$s$\ and\
$\bar{s}.$\  This, again, requires that the $s$\ and/or the $\bar{s}$ interact with the
remnants of the proton.

Thus both $\Delta s$ and $G^s$ give a measure of strange quarks in the proton which have 
some
interaction with the rest of the proton and hence are an integral part of the proton's
wavefunction.  (There is, however, a contribution to $\Delta s$ coming only from mass 
effects where
interactions in the proton are not necessary.)  Thus sizeable values of $\Delta s$ and $G^s$ 
would
indicate that strange quarks play an essential role in the proton's wavefunction.  If both 
$\Delta
s,$ after subtracting the anomaly, and $G^s$ turn out to be small it would strongly suggest 
that
the $s\bar{s}$ pairs in the proton are short-lived and play no essential role in 
understanding the
proton. These are, indeed, important and interesting issues.  It should be noted that there
already is significant evidence that the non-strange sea is strongly interacting in the proton
since the $\bar{u}/\bar{d}$ radio is x 
-dependent\cite{CCFR,Pro}.  However, it could well 
be that
strange quark fluctuations are significantly shorter lived and interact more weakly in the 
proton
than do the non-strange fluctuations.

\section{Heavy ion collisions and
small-x physics}
\subsection{Two main motivations}

There are at least two strong arguments for studing relativistic heavy ion collisions.  The 
first,
of course, is that such collisions offer the possibility of producing, at least temporarily, a 
new
state of matter, the deconfined quark-gluon plasma.  The second motivation is to study high 
field
strength QCD.  At the very early stages of a heavy ion collision, well before equilibration 
occurs,
very high values of the QCD field strength, $F_{\mu\nu}^a,$ are reached.   What the 
properties of
such a system are, and how large $F_{\mu\nu}^a$ can actually become, are fascinating 
questions
which theorists have only recently began to investigate.

\subsection{Lessons from QED?}

In quantum electrodynamics electric fields, having a coherence length and lifetime greater 
than
$1/m_e,$ have a maximum attainable value, $\vert \vec{E}\vert \sim {m_e^2\over e}.$  
Larger values
of $\vec{E}$ are immediately shielded by the copious creation of $e^+e^-$ pairs.  Thus, 
crudely
speaking, one can say that producing ${1\over m_e^4} \vec{E}^2 \geq 1/\alpha$ will 
result in a
breakdown of the QED vacuum.  What about a hot QED plasma?  How big are the field 
strengths in
such a system?  The question is relatively easy to answer.  Suppose we look over a region 
of
radius  $r_c$ and ask what is the size of the electric field coherent over a sphere that size.  
The
answer is given by counting the number of photons in the plasma having $\vert\vec{k}\vert 
\leq
1/r_c.$  Thus,

\begin{equation}
r_c^4 \vec{E}_c^2 \propto r_c^3 \int d^3k \Theta (1/r_c-\vert\vec{k}\vert){1\over
e^{\omega_{\vec{k}}/T}-1.}
\end{equation}

\noindent Since $[e^{\omega_{\vec{k}}/T}-1]^{-1}\approx T/\omega_{\vec{k}}$ for 
small
$\vert\vec{k}\vert$ it is clear that $r_c^4\vec{E}^2$ grows as $r_c$ grows and that the 
growth stops
when the plasma frequency $\omega_P \propto eT$ is reached.  Thus for $r_c \sim (eT)^{-
1}$

\begin{equation}
r_c^4\vec{E}_c^2 \propto 1/e
\end{equation}

\noindent and this is much less than $1/\alpha$ the maximum allowed value in the vacuum.  
The
value of the maximum field found in a plasma is lower than that of the vacuum because the 
plasma
has many electrons having momenta much greater than $eT$ which are effective in 
shielding electric
field fluctuations whose value is greater than that given by (23).

\subsection{Crude picture of early stages of a heavy ion collision}

Likely at RHIC energies and certainly by LHC energies semihard gluon production will 
dominate the
transverse energy freed in relativistic heavy ion collisions.  At the moment there is a
semiquantitative understanding of the early stages of a heavy ion collision coming from
semihard gluon production\cite{Bla,Lin,Esk,hep,Wan}.  Imagine an ion-ion collision in 
the center of
mass frame.  Just before the collision the wavefunction of each of the ions has many small-
x
gluons.  During the collision large numbers of these small-x gluons are freed by elastic
gluon-gluon scattering as illustrated in Fig.2.  The freed gluons which dominate the 
produced
transverse energy are expected to have $p_\perp \approx 1 GeV$ at RHIC and $p_\perp 
\approx 1-3
GeV$ at LHC.  At LHC the gluons are clearly within the hard scattering regime while at 
RHIC the
hard scattering contribution should give a reasonable estimate of the freed transverse energy
at early times after the collision.  Rough estimates indicate that\cite{Bla,Lin,Esk,hep}

\begin{equation}
{dE_\perp\over dy} \approx 1\  TeV {\rm at\  RHIC}
\end{equation}

\noindent and

\begin{equation}
{dE_\perp\over dy} \approx 15 TeV {\rm at\ LHC}
\end{equation}

\noindent with the 1\ TeV at RHIC coming from $10^3$ gluons and the 15 TeV at LHC 
coming from
$5x10^3$ gluons\cite{hep}

\begin{figure}[htb]
\epsfxsize=3in
\epsfysize=1.5in  
\epsfbox[0 0 297 105 ]{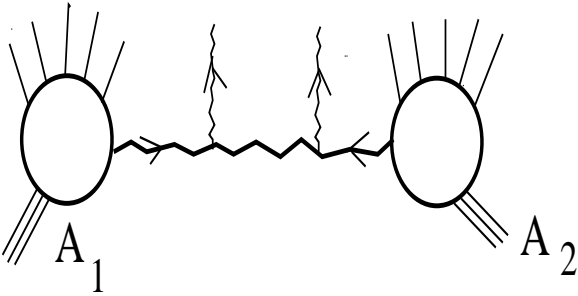} 
\caption{The dominant process in estimating gluon production in ion-ion collisions}
\label{Fig.2-928}
\end{figure}

\subsection{A closer look at the early times after an ion-ion collision}

The McLerran-Venugopalan model\cite{McL} is a nice framework within which to look 
more closely at
the early stages of a high energy heavy ion collision.  While the McLerran-Venugopalan 
model cannot
be expected to be correct in the details of a heavy ion collision it should be a reasonable 
guide
as to how the reaction proceeds.  One begins by looking at the distribution of valence 
quarks in a
high energy ion.  The valence quarks are found in a Lorentz-contracted longitudinal disc of 
size
$\Delta z = 2R\cdot{m\over p}$ where \ $m$\ is the nucleon mass and \ $p$\ the 
momentum per nucleon
of the ion.  For our purposes we consider $\Delta z = 0$ so that one can imagine the 
valence
quarks having a two-dimensional number density of quarks per unit area

\begin{equation}
n_q(b) = 6\rho{\sqrt{R^2-b^2}},
\end{equation}

\noindent where $\rho$ is the normal nuclear number density, which we assume to be 
constant in the
nucleus, while \ $b$\ is the impact parameter measured from the center of the nucleus in a 
direction
perpendicular to the direction of motion of the nucleus.  The situation is illustrated in Fig.3.

\begin{figure}[htb]
\epsfxsize=3.5in
\epsfysize=2.5in
\epsfbox[0 0 172 128]{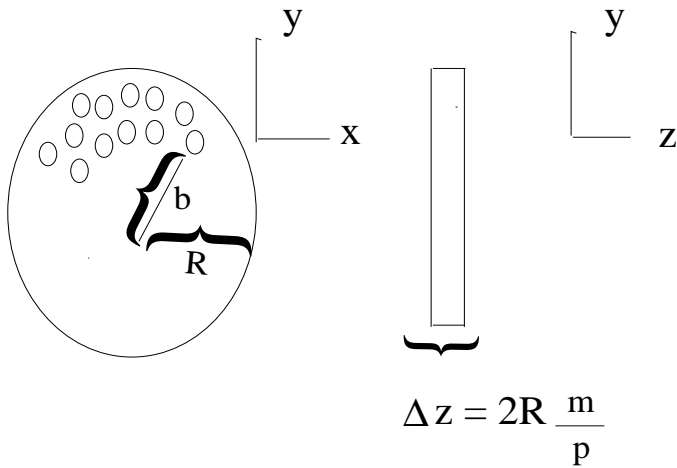}
\caption{The spatial distribution of valence quarks in a high energy heavy ion}
\label{Fig.3-928}
\end{figure}

The valence quark density, given by (26), is the source for soft gluons corresponding to 
the
Weizs\"acker-Williams field of  $n_q$\cite{Jal,Kov}.  The color charge at a given impact 
parameter
comes from a random addition of the color charges of each of the valence quark at that 
impact
parameter.  A single quark gives  gives ${\alpha C_F\over \pi} \ell n {Q^2\over \mu^2}$ 
gluons at
scale $Q^2$ per unit rapidity so that one expects the number density of gluons per unit area 
and
per unity rapidity to be

\begin{equation}
n_g(b,Q) = n_q(b) \cdot {\alpha C_F\over \pi} \ell n Q^2/\mu^2
\end{equation}

\noindent in the ``additive'' Weizs\"acker-Williams approximation.  (If $\mu \approx 100 
MeV$ (27) 
works reasonably well for a single proton.)  However, a more careful application of the
non-Abelian Weizs\"acker-Williams approximation shows that there can never be more than 
$1/\alpha$
gluons occupying the same transverse area\cite{Jal,Mue}.  Thus

\begin{equation}
n_g^{max}(b,Q) = {N_c^2-1\over \pi\alpha N_c} \cdot {1\over \pi(2/Q)^2} = {(N_c^2-
1)Q^2\over
4\pi^2\alpha N_c}
\end{equation}

\noindent where we take the area of a gluon at scale\  $Q$\  to be $4\pi/Q^2$  and we have
inserted the color factors to agree with the results of Refs.30 and 32. Thus so long as (27) 
is less
than (28) it should be a reasonable picture of the gluon distribution in a large nucleus.  
When
$Q^2$ is small enough so that (27) is greater than (28) one says that the gluon distribution 
has
saturated\cite{Gri} with
$n_g^{max}$ being the saturated distribution.  We can make (27) a little less model 
dependent by
identifying
${3\alpha C_F\over
\pi}\ \ell n Q^2/\mu^2$ with
$xG(x,Q^2)$ the gluon distribution in a nucleon.  In that case (27) becomes, using (26),

\begin{equation}
n_g(b,Q) = 2\rho{\sqrt{R^2-b^2}}\  x G(x,Q^2).
\end{equation}

\noindent Equating (28) and (29) gives

\begin{equation}
Q_{sat}^2 = 8\pi^2{N_c\over N_c^2-1} \alpha\rho{\sqrt{R^2-b^2}}\  x G(x,Q_{sat}^2)
\end{equation}

\noindent as the saturation momentum.  For $Q^2$ values below $Q_{sat}^2$ (28) should 
be the gluon
distribution in the nucleus, while for $Q^2 > Q_{sat}^2$ (29) should be appropriate.  The
$Q_{sat}^2$ value given by (30) at $b=0$ is very close to that given long ago in Ref.24, 
differing
only by a factor $3/4$ and is the same as given in Refs. 30 and 32 though the
discussion here has been much simplified.  It is likely that the value of $Q_{sat}^2$ at 
RHIC is
about ${1\over 2} - 1 GeV^2$ or so while the value at LHC may well be in the 
$Q_{sat}^2 \approx 2
GeV^2$ region at $b=0.$

There is pretty good control of the ion's wavefunction in the McLerran-Venugopalan 
model, however,
so far no one has succeeded in doing a realistic calculation of the ``scattering'' to determine
which gluons are freed at early times. One would guess that all gluons having transverse 
momentum
below  $Q_{sat}$ would be freed while not so many of the gluons above $Q_{sat}$ 
become free. A
good calculation of the gluon interactions to determine exactly which gluons are freed is a 
key 
calculation for progress in understanding the early stages of heavy ion collisions.

\subsection{What about the proton wavefunction at small x?}

Is the picture of the small-x gluon distribution in a nucleus  unique to large nuclei or can the
same picture apply to the proton?  Saturation, or a maximum field strength, comes about 
when
many gluons are available to occupy the same region of phase space in a wavefunction. In a 
large
nucleus the source producing those gluons is the large  number of valence quarks.  In a 
proton a
similar phenomena can occur at extremely small values of x, where now the large rapidity 
interval,
$y=\ell n\  1/x,$ available for gluon evolution can lead to high gluon number densities. 
Theoretical control over this very small-x regime is not perfect yet, and the importance of 
BFKL
or x-evolution is not yet clear in the HERA regime, so one must turn to phenomenology to 
see if
there is evidence for saturation effects.  The answer is, perhaps, there is some evidence 
from
low-$Q^2$ very small-x structure function data at HERA.

\begin{figure}[htb]
\epsfxsize=3.5in
\epsfysize=2.5in
\epsfbox[0 0 539 252]{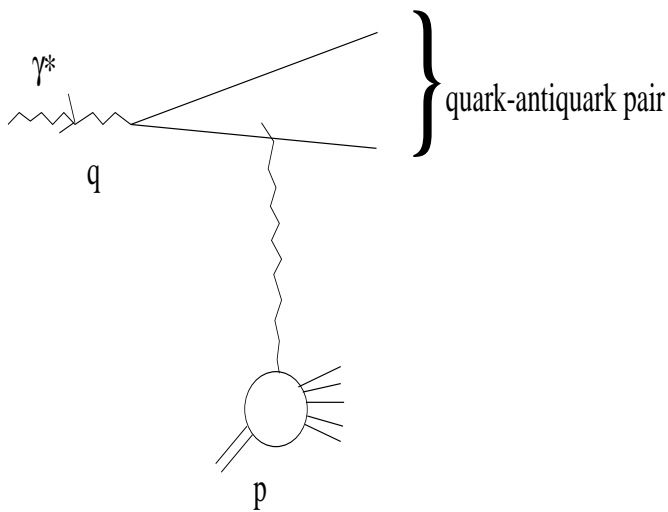}
\caption{Viewing the structure function of the proton in terms of the scattering of a
high energy quark-antiquark pair.}
\label{Fig.4-928}
\end{figure}

Let me briefly describe a useful way to view structure functions for the purposes of seeing 
if
saturation effects are present.  It is convenient to choose a collinear frame where the proton
momentum, P, and the virtual photon momentum, $q,$  take the form

\begin{eqnarray}
P&\approx& (P + {m^2\over 2P}, 0,0,P)\\
\nonumber q&\approx& ({\sqrt{q^2-Q^2}},0,0,-q)
\end{eqnarray}

\vspace{1cm}

\begin{figure}[htb]
\epsfxsize=3.5in
\epsfysize=2.5in
\epsfbox[46 186 497 673]{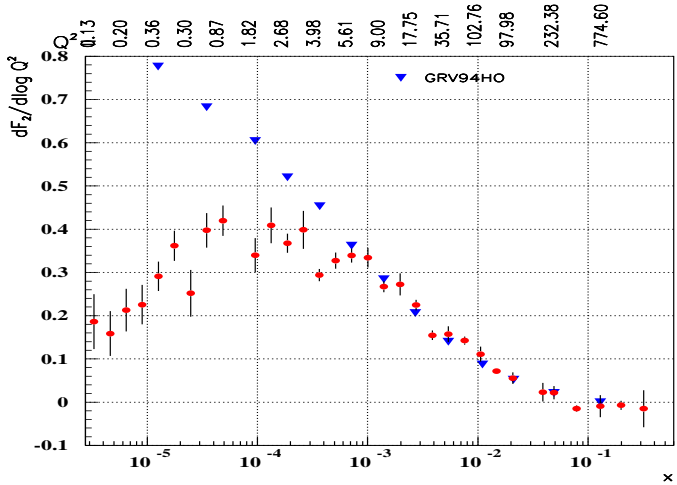}
\caption{The Caldwell plot of HERA data.}
\label{Fig.5-928}
\end{figure}
\vspace{1cm} 

\begin{figure}[htb]
\epsfxsize=3.5in
\epsfysize=2.5in
\epsfbox[46 186 502 676]{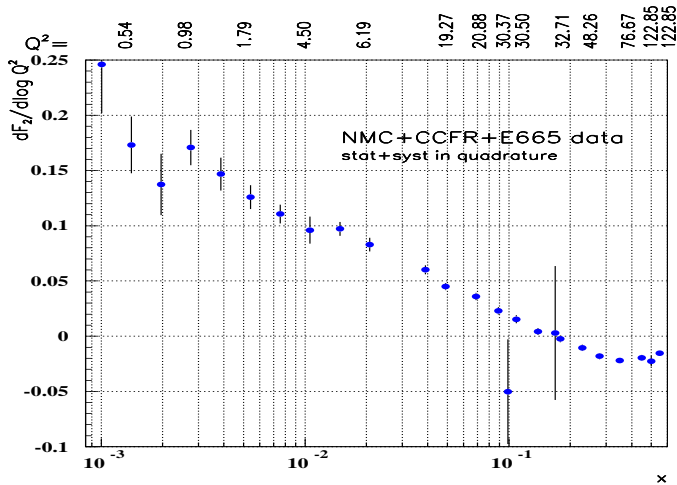}
\caption{Caldwell's compilation of points from fixed target experiments}
\label{Fig.6-928}
\end{figure}

\noindent with $P/q$  very large and $q/Q$ large  but fixed independently of $1/x = 
{2P\cdot q\over
Q^2.}$  Thus all
$x-evolution$ is included in the proton's wavefunction and the scattering can be viewed as
illustrated in Fig.4.  Before the collision the virtual photon splits into a quark-antiquark pair
which then interacts with the gluon field of the proton to produce an inelastic collision.   A.
Caldwell\cite{Cal} suggested looking at $Q^2{\partial F_2(x,Q^2)\over \partial Q^2}$ as a 
useful way
of seeing very small-x effects at moderate $Q^2.$  This is useful because

\begin{equation}
Q^2{\partial F_2(x,Q^2)\over \partial Q^2} \propto Q^2 \sigma_{{q}\bar{q}}(\Delta 
x_\perp \sim 1/Q)
\end{equation}

\noindent where $\sigma_{{q}\bar{q}}$ is the cross section of the ${q}\bar{q}$ pair to 
interact
with the proton. The transverse coordinate separation of the pair is proportional to $1/Q.$  
If
gluons are reasonably dilute in the proton, so that the parton picture applies, one has

\begin{equation}
\sigma_{{q}\bar{q}} \propto {\alpha\over Q^2}\  x G(x,Q^2)
\end{equation}

\noindent with \ $G$\ the gluon distribution in the proton.  The  $1/Q^2$ on the right-hand 
side of
(33) reflects the diluteness of gluons 
in the proton.  On the other hand, if gluons are packed
densely in the proton one reaches the unitarity limit and cross sections become geometric in 
which
case

\begin{equation}
\sigma_{{q}\bar{q}}\  \propto \ \pi R_0^2
\end{equation}

\noindent where $R_0$ is the radius of these proton. Using (33) and (34) in (32) we see 
that a
signal for saturation is that $Q^2 {\partial F_2\over \partial Q^2}$ should grow linearly in 
$Q^2.$ 
HERA data is shown in Fig.5 while fixed target data is shown in Fig.6.  Data are plotted as 
a
function of \ $x$\ on the lower axis and as a function of\  $Q^2$\ on the upper axis.  The 
turnover
in the HERA data at \
$Q^2 \approx 2,$\ and the lack of such a  turnover in the larger \ $x$\ fixed target data may 
be
the indication of parton saturation for all transverse momenta less than about $1-1\ 1/2
GeV$\cite{Got,DIS,Bier}.

\subsection{Why is equilibration nontrivial?}

We have seen that there are a great many gluons in the wavefunction of a high energy 
heavy ion and
that many of these gluons are likely freed in a central ion-ion collision.  With the large 
gluon
densities present it would seem to be a simple matter to reach equilibration.  However, the 
gluons
are located in a single layer, perpendicular to the axis of collision.  Thus immediately after 
the
collision the gluon density will begin to rapidly decrease and it is not certain, a priori, 
whether
equilibration will set in before the system falls apart. There are some pretty good 
indications from
Monte Carlo calculations\cite{Gei,Zha,Won} that kinetic equilibration does indeed occur at 
RHIC and
LHC energies, however, it would be nice to see  more clearly, in analytic calculation what 
the
issues are and how certain it is that equilibration occurs.  In what follows I give a very
simplified version of a criterion for equilibration in the McLerran-Venugopalan model.

We imagine a head-on collision of two heavy ions each of which has gluons in its 
wavefunction
saturated up to a scale\  $Q.$\  We suppose that during a time on the order of $1/Q$ the 
gluons
having $k_\perp \leq Q,$ in the central unit of rapidity are freed while those having 
$k_\perp
\geq Q$ are virtual fluctuations only and are not freed.  Then follow a particular freed gluon
in the central unit of rapidity.  (We may suppose that $k_\perp$ of the gluon is $k_\perp
\approx Q,$ since there is little phase space for
$k_\perp << Q$\ and freed gluons having $k_\perp >> Q$\ are few.)  If this gluon has a 
collision
with momentum transfer on the order of  $Q$  it has gone a significant way toward 
equilibration. 
Thus as an equilibration criterion we take    

\begin{equation} 
1 = \int_{\tau_0}^\tau {dt\over \lambda (t)},
\end{equation}

\noindent where $\tau_0=1/Q, \lambda(t)$ is the mean free path and $\tau$ is some time 
which should
obey $\tau \leq R.$  (If the gluon does not experience a hard scattering in $\tau \leq R$ it is
unlikely that a later scattering will occur because the expansion becomes 3-dimensional 
rather than
1-dimensional.)  Now

\begin{equation}
{1\over \lambda(t)} = \rho_g(t) \sigma_{gg\to gg}
\end{equation}

\noindent where

\begin{equation}
\rho_g\cdot\ g(t) = {n_g^{max}\over t} = {(N_c^2-1)Q^2\over 4\pi^2\alpha N_c t}
\end{equation}

\noindent and

\begin{equation}
\sigma_{gg\to gg} = \bigl({\alpha N_c\over \pi}\bigr)^2 {4\pi^3\over (N_c^2-1)Q^2}
\end{equation}

\noindent In using (37) we neglect the b-dependence of the saturation momentum.  In (38) 
we have
integrated ${d\sigma\over dt} = ({\alpha N_c\over \pi})^3 {4\pi^3\over (N_c^2-1)t^2}$ 
over $t$ from
$-t = Q^2$ to $\infty .\  \alpha$ in (37) and (38) should be evaluated at $Q^2.$  Thus (35) 
becomes

\begin{equation}
1 = {\alpha N_c\over \pi}   \ell n\  \tau/\tau_0
\end{equation}

\noindent or, using $\alpha(Q^2) = {1\over b\ell n Q^2/\wedge^2}$ with $b = {11 N_c-
2N_f\over
12\pi},$

\begin{equation}
\tau = {1\over \wedge} ({Q\over \wedge})^{2\pi\ b/N_c-1}.
\end{equation}

\noindent Since $Q^2 \propto R$ from (30) equilibration will occur if

\begin{equation}
{\pi \ b\over N_c} - {1\over 2}\  <\  1
\end{equation}

\noindent which is the case.  It is reassuring to see that the
equilibration seems to be met in the McLereran-Venugopalan model, 
but it is , perhaps, a little disturbing that it seems to be accidental and not a fundamental
property of high energy heavy ion collisions.

\section{Perturbative versus nonperturbative QCD}

Full nonperturbative calculations of physical quantities are desirable but hard to get in 
QCD.  In
hard probes of quark and gluon structure in hadrons the operator product expansion 
separates hard
from soft physics where the hard part is calculable perturbatively and the soft part can be
parametrized in a general way.  The operator product expansion takes the form

\begin{equation}
\int e^{iqx} d^4x\  j(x) j(0) \mathrel{\longrightarrow\atop q^2 large} {\cal{O}}_1E_1(\alpha(Q)) 
+ {1\over
Q^2}
{\cal{O}_2 E_2(\alpha(Q)) + \cdot\cdot\cdot}
\end{equation}

\noindent where ${\cal{O}}_1$ is the leading twist term and ${\cal{O}}_2$ and the next-
to-leading
twist term with $E_1$ and $E_2$ coefficient functions.  An interesting subtlety arises 
because $E_1$
has a divergent perturbation series

\begin{eqnarray}
E_1(\alpha)&=& \sum_n E_{1n} \alpha^n\\
\nonumber E_{1n}{\sim\atop n\ large} &e_1& n! \beta_2^n n^\gamma (1+0({1\over 
n})),
\end{eqnarray}
 ...

\noindent where $\beta_2$ is the first coefficient of the $\beta-$function.  The ambiguity in 
the
sum (40) due to the divergent perturbation series is of size $1/Q^2$ and means that the 
separation
of the ${\cal{O}}_1$ and

$t{O}_2$ terms in (39) is not without 
ambiguity\cite{ler}.

For example, for the $F_3$ structure function in neutrino proton scattering one commonly 
writes

\begin{equation}
xF_3(x,Q^2) = xF_3^{LT}(x,Q^2) + {h(x)\over Q^2}
\end{equation}

\noindent with $F_3^{LT}$ the leading twist contribution and the $h/Q^2$ the higher twist 
term. 
Recently an interesting observation has been made by Kataev\cite{Kat} and his 
collaborators that
$h(x)$ depends on the level to which the perturbation theory analysis of $F_3^{LT}$ is 
done.  If
$F_3^{LT}$ is evaluated in a leading order renormalization group formalism a large 
$h(x)$ emerges
from the fit. If a next-to-leading order formalism issued for $F_3^{LT}$ a somewhat 
smaller but
still substantial contribution from $h$ is needed.  When $F_3^{LT}$ is evaluated in a
next-to-next-to-leading formalism very little room is left for $h.$  This is shown in Figs.7-
9
taken from Ref.9.  I think there is an important lesson here.  It is not clear how to separate
leading and higher twist terms and the separation will depend on the level to which the
perturbative calculation is done.

\begin{figure}[htb]
\epsfxsize=3.5in
\epsfysize=2.5in
\epsfbox[34 140 577 652]{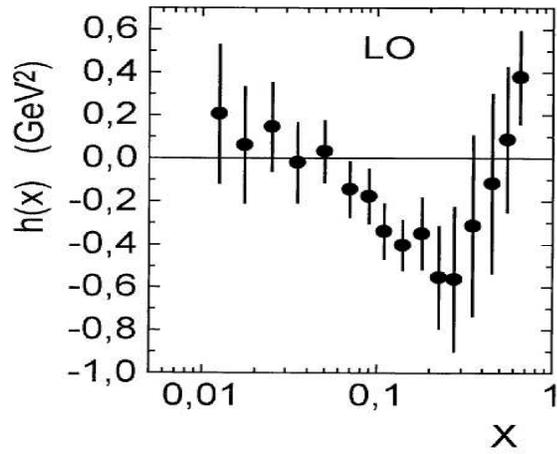}
\caption{The results of the LO extraction of the x-shape of the
 twist-4 contribution  h(x)}
\label{Fig.7-928}
\end{figure}

\begin{figure}[htb]
\epsfxsize=3.5in
\epsfysize=2.5in
\epsfbox[34 -90 832 653]{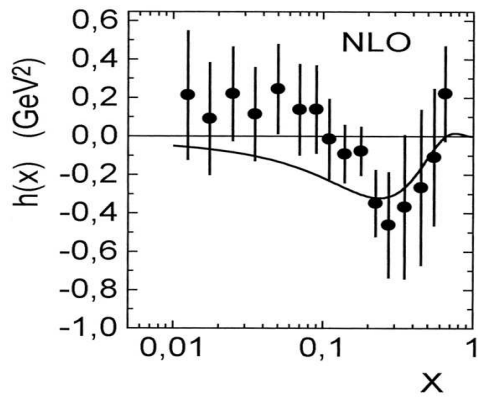}
\caption{The results of the NLO extraction of the x-shape of the twist-4 contribution 
h(x).  For comparison, the IRR-model prediction of Ref.[22], obtained using the NLO MRS 
parametrization, is
also depicted.}
\label{Fig.8-928}
\end{figure}

\begin{figure}[htb]
\epsfxsize=3.5in
\epsfysize=2.5in
\epsfbox[0 0 330 294]{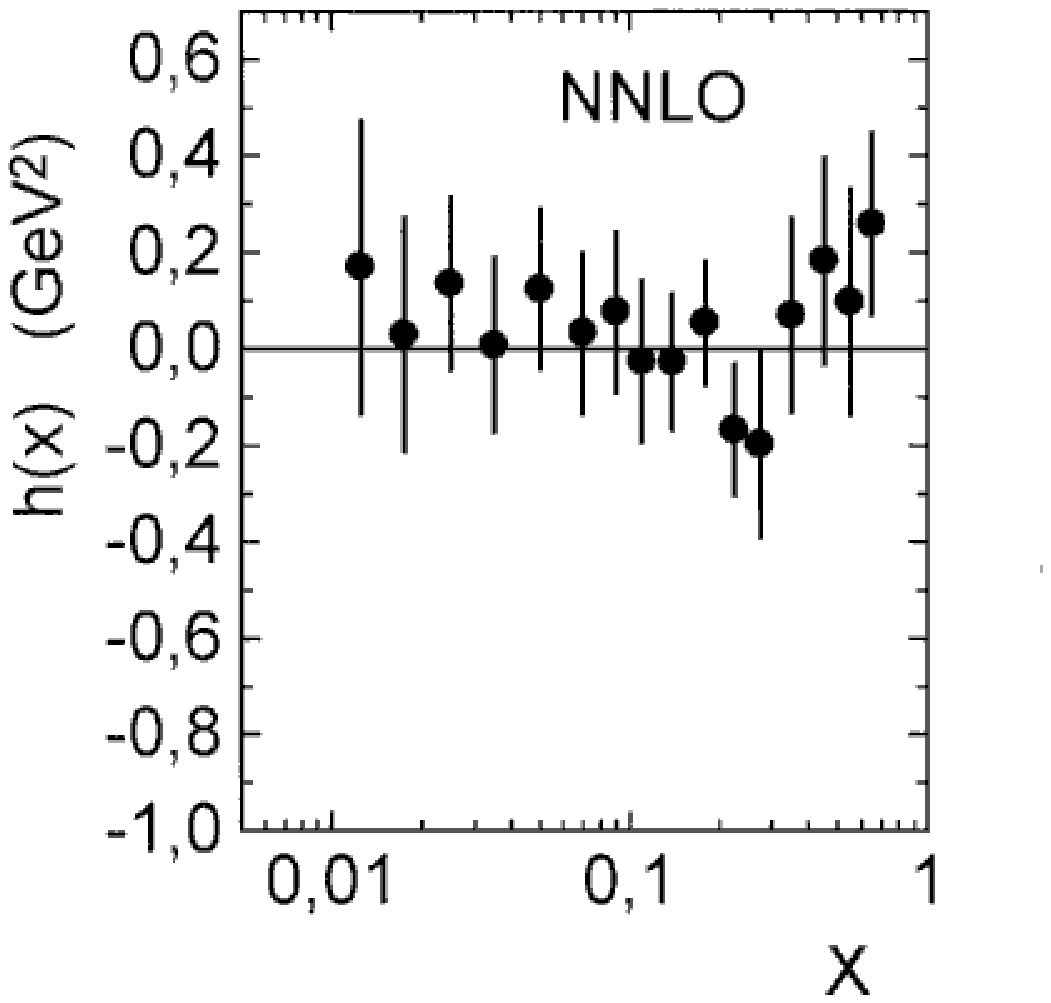}
\caption{The results of the NNLO extraction of the x-shape of the twist-4 contribution 
$h(x).$}
\label{Fig.9-928}
\end{figure}

\end{document}